\documentclass{revtex4}

\usepackage{amsmath,latexsym,amssymb,verbatim,enumerate,graphicx,amsfonts}
\usepackage{mathrsfs,hyperref,amsthm,braket,dsfont}

\usepackage{bm}
\usepackage{tikz}
\usetikzlibrary{positioning,shapes}
\usepackage{float}
\usepackage{color}

\theoremstyle{plain}
\newtheorem{theo}{Theorem}[section]

\newtheorem{rem}[theo]{Remark}

\begin{document}

\title{Adversarial vs cooperative quantum estimation}

\author{Milajiguli Rexiti\footnote{Corresponding author. Electronic address: milajiguli.milajiguli@unicam.it}}
\affiliation{Department of Mathematics and Physics, Xinjiang Agricultural University, 830053 Urumqi, China\\School of Advanced Studies, University of Camerino, 62032 Camerino, Italy}
\author{Stefano Mancini}
\affiliation{School of Sience and Technology, University of Camerino, 62032 Camerino, Italy\\
INFN-Sezione di Perugia, I-06123 Perugia, Italy} 

\begin{abstract}
We address the estimation of a one-parameter family of isometries taking one input into two output systems. This primarily allows us to consider imperfect estimation by accessing only one output system, i.e. through a quantum channel.
Then, on the one hand, we consider separate and adversarial control of the two output systems to introduce the concept of \emph{privacy of estimation}. On the other hand we conceive the possibility of  
separate but cooperative control of the two output systems.
Optimal estimation strategies are found according to the minimum mean square error. This also implies the generalization of Personik's theorem to the case of local measurements. 
Finally, applications to two-qubit unitaries (with one qubit in a fixed input state) are discussed.
\end{abstract}
\date{\today}
\maketitle


\section{Introduction}

Parameter estimation, which plays a central role in mathematical statistics, becomes of tantamount importance in quantum information processing too (see e.g. \cite{Paris}). A paradigmatic example is represented by the estimation of a parameter characterizing quantum states transformations \cite{B05,H06,K07}. These are ideally unitary transformations, however, in practice one has to deal with noisy quantum maps, hence parameter estimation has been extended to quantum channels \cite{SBB02,FI03,Zetal06}. 

Several aspects of quantum channels have been investigated in recent years. One of this is the \emph{privacy}, i.e. the amount of information that traverses a channel without being intelligible to a third party besides the legitimate sender and receiver \cite{CWY04}.
Its determination amounts to consider a \emph{competition} between the receiver, as accessing the output channel information, and a third malicious party, as accessing the information lost into the environment.
However recently also \emph{cooperation} between actors of a quantum communication set up is receiving an increasing attention (see e.g. \cite{BN14,Ketal16}).
Our aim here is to analyze these two opposite strategies in the quantum estimation theory framework.

It is well known that any quantum channel, being a completely positive and trace preserving map, admits an isometry as dilation \cite{Stinespring}.
Hence we can conceive the estimation of a family of isometries \emph{through} quantum channels.
More precisely, given a one parameter family of isometries $\{V_{s}^{A\to BF}\}$,  
we consider the parameter $s$'s estimation by accessing only the system $B$. 
This amounts to use the quantum channel between $A$ and $B$
of which $V_{s}^{A\to BF}$ represents the Stinespring dilation \cite{Stinespring}.
Estimation through such a channel basically models a realistic situation where not all output information can be gathered.

In this context we shall first consider the system $F$ under control of a malicious being. 
Then the question arises of what are the conditions under which a legitimate user controlling 
the $B$ system (besides $A$ ones) can perform a better estimation. 
Second, we shall consider the system $F$ under control of a benevolent helper. Then the question arises of what would be the advantage in estimating locally, but cooperatively the isometrie (i.e. with local measurements and classical communication). 

We shall address these issues by considering the mean square error as figure of merit and
pursuing its minimization. 
In one case we introduce the concept of \emph{private estimation}, which is defined as the difference  of the mean square error of the $F$ system and the $B$ system (Section \ref{sec:aqe}).
In the other case the possible measurements are constrained to be \emph{local} in systems $B$ and $F$, thus
we generalize the Personik's theorem \cite{Pers71}, which represents the standard way to 
minimize the  mean square error (Section \ref{sec:cqe}).
Finally, the effectiveness of these approaches in  estimation is shown 
with applications to two-qubit unitaries, 
regarded as isometries by fixing one qubit input state
(Section \ref{sec:application}).


\section{Adversarial quantum estimation}\label{sec:aqe}

We start with studying the situation that two parties compete in estimating the same parameter and we want to figure out when one party, considered legitimate, can outperform the other.
To formalize this let us take a family of isometries
\begin{equation}
V_{s} : \mathscr{H}_A \rightarrow \mathscr{H}_B \otimes \mathscr{H}_F,
\end{equation}
parametrized by ${s} \in \mathscr{I}\subset\mathbb{R}$.
The parameter $s$ is assumed to have an a priori probability distribution function $p(s)$ 
over $\mathscr{I}$.
Furthermore, we consider $A$ as the probe system prepared in the state $\rho_A$.
Then the output on $B$ reads
\begin{equation}
\rho_B({s})={\rm Tr}_F\left( V_{s} \rho_A V_{s}^\dag\right)=:{\cal N}(\rho_A).
\label{calN}
\end{equation}
On this state we perform a measurement whose outcome provides an estimate of the unknown parameter ${s}$. 

The goodness of this process can be measured by 
the average quadratic cost function corresponding to the mean square error
\begin{equation}
\bar{C}^B:=\int_\mathscr{I} p({s}) {\rm Tr}\left[\rho_B ({s}) \left(\hat S_B-{s} I \right)^2\right]d{s},
\label{costB}
\end{equation} 
where $\hat S_B $ is the measurement operator that we use to estimate $s$.
The best of such operator is obtained by minimizing $\bar{C}^B$.
Personik's theorem \cite{Pers71} states that the minimum mean square error estimator must satisfy the following (linear) equation
\begin{equation}\label{Pers}
W^{(0)}_B\hat{S}_B+\hat{S}_BW^{(0)}_B=2W^{(1)}_B,
\end{equation}
where
\begin{subequations}
\begin{eqnarray}
{W}^{(0)}_B&:=& \int_\mathscr{I} p({s}) {\rho}_B({s}) d{s}, \\
{W}^{(1)}_B&:=& \int_\mathscr{I} {s} \, p({s}) {\rho}_B({s}) d{s}.
\end{eqnarray}
\label{Wini}
\end{subequations}
Whether the solution of \eqref{costB} will result in a biased or unbiased estimator
will depend on the explicit form of the $W$s in \eqref{Wini}, however this is not relevant for the following.
Instead it is worth noticing that by using the spectral decomposition 
$\hat{S}_B=\int \hat{s}_B  \,d \Pi(\hat{s}_B)$, 
with the Probability Operator Valued Measure (POVM) $d\Pi(\hat{s}_B)
:=|\hat{s}_B\rangle\langle\hat{s}_B|\, d\hat{s}_B$
defined by the eigenvectors of $\hat{S}_B$, Eq.\eqref{costB} can be recast in the form
\begin{eqnarray}
\bar{C}^B=\int_\mathscr{I} \int_\mathscr{I} p(s) \left(\hat{s}_B  -{s}\right)^2{\rm Tr}
\left[\rho_B (s)  \Pi(\hat{s}_B)  \right]ds \,d \hat{s}_B,
\end{eqnarray}
where ${\rm Tr}\left[\rho_B (s)  \Pi(\hat{s}_B)  \right]$ represents the conditional probability of getting the estimate value $\hat{s}_B$ given the prameter value $s$.

Although the above approach seems limited to projective measurements, it can be shown 
that when a single parameter $s$ has to be estimated with minimum mean square error,
the POVM defined by the eigenvectors of the operator $\hat{S}_B$ satisfying \eqref{Pers} represents the optimum estimation strategy 
overall possible POVMs \cite{Helstrom}.

On the other hand we can consider the state emerging from the channel complementary to $\cal N$ 
in Eq.\eqref{calN}, namely 
\begin{equation}
\rho_F({s})={\rm Tr}_B\left(V_{s} \rho_A V_{s}^\dag\right)=:{\widetilde{\cal N}}(\rho).
\label{tildecalN}
\end{equation}
If this is controlled by an adversary, a strategy similar to the above can be employed to estimate 
${s}$  and leads to $\bar{C}^F_{min}$ with a suitable optimal measurement $\hat S_F$.

By considering the system $B$ (as well as $A$) hold by a legitimate user,
we define the \emph{privacy of estimation} through the difference between the minimum of the average quadratic cost functions 
\begin{equation}\label{Pe}
{\cal P}_e:=\max\left\{\bar{C}^{F}_{min}-\bar{C}^{B}_{min},0\right\}.
\end{equation}
Whenever it results positive it means that $\bar{C}^{B}_{min}<\bar{C}^{F}_{min}$ and hence $B$ can estimate $s$ better than $F$.
This definition of privacy assumes that the adversary can control the system $F$ and at the same time has information about the input state. A weaker notion of privacy can be introduced by assuming the adversary with no information about the input state. This amounts to consider $\bar{C}^F_{min}$ in \eqref{Pe} averaged overall possible input states.


\section{Cooperative quantum estimation}\label{sec:cqe}

Here, in contrast to Section \ref{sec:aqe}, we analyze the possibility that two parties cooperate while trying to estimate the same parameter and we ask when this is advantageous. More specifically, 
suppose now that $B$ and $F$ are not adversary, but they want to cooperate in order to estimate 
$s$, though acting locally. Starting from $\rho_A$ we find the joint output on $B$, $F$ as
\begin{equation}
\rho({s})=V_{s} \rho_A V^\dag_{s}.
\end{equation}
Then local measurements are performed on this state and the outcomes (after classical communication) provide an estimate of the unknown parameter $s$. 

We want to find the optimal local measurement operator $\hat S = \hat S_B \otimes\hat S_F $
(with  $\hat S_B$, $\hat S_F$ hermitian operators in $\mathscr{H}_B$, $\mathscr{H} _F$ respectively),  
 such that the average quadratic cost function, corresponding to the mean square error
 \begin{equation}
\bar{C}:=\int_\mathscr{I} p({s}) {\rm Tr}\left[\rho ({s}) \left(\hat S-{s} I \right)^2\right]d{s},
\label{cost}
\end{equation} 
is minimum.

For the sake of convenience we define:
\begin{subequations}
\begin{eqnarray}
{W}^{(0)}&:=&   \int_\mathscr{I} p({s}) {\rho}({s}) d{s}, \\
{W}^{(1)}&:=&    \int_\mathscr{I} {s} \, p({s}) {\rho}({s}) d{s}.
\end{eqnarray}
\label{W}
\end{subequations}
Then the optimal local measurement can be found according to the following Theorem.

\begin{theo}\label{P2}  
The optimal local measurement $ \hat S= \hat S_B \otimes\hat S_F $, with 
$ \hat S_B$, $\hat S_F$ hermitian operators in $\mathscr{H}_B$, $\mathscr{H} _F$ respectively,
satisfy the following set of coupled equations:
\begin{subequations}
\begin{eqnarray}
{\widetilde W}^{(0)}_B \hat S_B +\hat S_B  {\widetilde W}^{(0)}_B =2  W^{(1)}_B,
\label{1a}  \\ 
{\widetilde W}^{(0)}_F \hat S_F +\hat S_F   {\widetilde W}^{(0)}_F=2  W^{(1)}_F,
\label{1b}
\end{eqnarray}
\label{1}
\end{subequations}
where 
\begin{subequations}
\begin{align}
{\widetilde W}^{(0)}_B &:={\rm Tr}_F \left \{W^{(0)}  \left(I_B \otimes\hat S_F\right) \right\}, \\
{\widetilde W}^{(0)}_F &:={\rm Tr}_B \left \{W^{(0)}  \left(\hat S_B \otimes I_F\right) \right\}, 
\end{align}
\label{WBF}
\end{subequations}
while, likewise Section \ref{sec:aqe}, 
$W^{(1)}_B= \int_\mathscr{I} {s} \, p({s}) \rho_B({s}) d{s}={\rm Tr}_F W^{(1)}$ and
$W^{(1)}_F = \int_\mathscr{I} {s} \, p({s}) \rho_F({s}) d{s}={\rm Tr}_B W^{(1)}$.
\end{theo}

\bigskip
The proof of this Theorem is reported in Appendix \ref{appA}.
\bigskip

\begin{rem}
Eqs.\eqref{1} is a set on nonlinear equations. In fact, if we first consider \eqref{1a} as a linear equation with respect to $\hat S_B$ and solve it, 
given that ${\widetilde W}^{(0)}_B$, ${\widetilde W}^{(0)}_B$ depend on 
$\hat S_F$, we will have a solution $\hat S_B\left (\hat S_F\right)$ depending on $\hat S_F$.
In turn this determines ${\widetilde W}^{(0)}_F$, ${\widetilde W}^{(0)}_F$ depending on 
$\hat S_F$ in \eqref{1b}. Thus the latter becomes a nonlinear equation whose solution is generally hard to find.  Eqs.\eqref{1} reduce to linear and uncoupled equations in case ${\widetilde W}^{(0)}$ and ${\widetilde W}^{(1)}$ are local. 
\end{rem}

Often isometries in Stinespring dilation are considered to be written as
\begin{equation}
V_s=U_s |0\rangle_E, 
\label{VU}
\end{equation}
where $U_s : \mathscr{H}_{A}\otimes \mathscr{H}_{E}\to\mathscr{H}_{B}\otimes \mathscr{H}_{F}$ are unitaries ($\mathscr{H}_A \sim \mathscr{H}_B$ and $\mathscr{H}_E \sim \mathscr{H}_F$).

\begin{rem}
${\widetilde W}^{(0)}$ and ${\widetilde W}^{(1)}$ in \eqref{WBF} result local in the case of non entangling $U_s$, and in particular when $U_s$ is simply the product $u_s \otimes u_s$, with $u_s : \mathscr{H}_A \rightarrow \mathscr{H}_B$ and  $ \mathscr{H}_A \sim \mathscr{H}_E$, we recover the original Personik's Theorem \cite{Pers71}.
\end{rem}


\section{Applications to two-qubit unitaries}\label{sec:application}

{ In this Section, following \eqref{VU}, we will apply the developed adversarial and cooperative estimation strategies to two-qubit unitaries $U_s : \mathbb{C}^2\otimes \mathbb{C}^2 \to \mathbb{C}^2 \otimes \mathbb{C}^2$. Of course it is meaningful to consider entangling unitaries.

The state in the system $A$ (probe's state) will be generically taken as 
\begin{equation}
\rho_A=\left(\sqrt{\gamma}\ket 0 +e^{i\varphi} \sqrt{1-\gamma}\ket 1\right)
\left(\sqrt{\gamma}\bra 0 +e^{-i\varphi} \sqrt{1-\gamma}\bra 1\right),
\label{rhoA2qubit}
\end{equation}
with $\gamma\in[0,1]$ and $\varphi\in[0,2\pi]$.}


{
\subsection{Phase damping channel dilation}\label{sec:app1}
}

{
An interesting example to start with is provided by 
\begin{equation}\label{UcR}
U_s=|0\rangle\langle 0|\otimes I +|1\rangle\langle 1|\otimes \left((\cos s) \, I+i(\sin s) \,\sigma_y\right),
\end{equation}
that describes a controlled rotation by an angle $s$. 
Following the standard convention we use $\sigma_x, \sigma_y, \sigma_z$ to denote the Pauli operators. 
The parameter space is
\begin{equation}\label{spaceS}
\mathscr{S}=\left\{ s: \frac{\pi}{2}\geq s  \geq 0\right\}.
\end{equation}
Eq.\eqref{UcR} represents the Stinespring dilation of the phase damping channel whose action between system $A$ and $B$, by referring to \eqref{VU}, is:
\begin{equation}\label{rhoBpdamp}
\rho_B={\cal N}(\rho_A)= {\rm Tr}_F\left(U_s \rho_A \otimes |0\rangle_E\langle 0| U_s^\dag\right) =K_0\rho_A K_0^\dag+K_1\rho_A K_1^\dag,
\end{equation}
with 
\begin{subequations}
\begin{align}
K_0&=|0\rangle\langle 0|+(\cos s)\, |1\rangle\langle 1|, \\
K_1&= -(\sin s)\, |1\rangle\langle 1|.
\end{align}
\end{subequations}
The effect of ${\cal N}$ is to attenuate the off diagonal matrix elements (with respect to the canonical basis $\{|0\rangle, |1\rangle\}$) of $\rho_A$ by a factor $(\cos s)$.

In turn, the output of the complementary channel is given by
\begin{equation}\label{rhoFpdamp}
\rho_F=\widetilde{\cal N}(\rho_A)= {\rm Tr}_B\left(U_s \rho_A \otimes |0\rangle_E\langle 0| U_s^\dag\right) =\widetilde{K}_0\rho_A \widetilde{K}_0^\dag+\widetilde{K}_1\rho_A 
\widetilde{K}_1^\dag,
\end{equation}
with 
\begin{subequations}\label{tildeKpdamp}
\begin{align}
\widetilde{K}_0&=|0\rangle\langle 0|, \\
\widetilde{K}_1&=\left((\cos s) I+i(\sin s) \sigma_y\right) |0\rangle\langle 1|.
\end{align}
\end{subequations}

Following the arguments of Sec.\ref{sec:aqe}, and using \eqref{rhoA2qubit}, we can readily compute the minimum cost function for \eqref{rhoBpdamp}
\begin{equation}\label{CBpdamp}
\bar{C}_{min}^B=\frac{\pi^2(\pi+2)-48\gamma(1-\gamma)(\pi-2)}{48(\pi+2)},
\end{equation}
as well as for \eqref{rhoFpdamp}
\begin{equation}\label{CFpdamp}
\bar{C}_{min}^F=\frac{\left(48\pi^2+4\pi^4-192(1-\gamma)^2\right)(1-\gamma)-\pi^6(1+\gamma)}{48\pi^2
\left(4-\pi^2-4\gamma-\pi^2\gamma\right)}.
\end{equation}
We may notice that they are independent on $\varphi$, while their dependence on $\gamma$ is shown in Fig.\ref{figpdamp}.

The minimum of $\bar{C}_{min}^B$ is achieved for $\gamma=\frac{1}{2}$ as one would expect due to the fact that an equally weighted superposition of canonical basis state vectors gives rise to largest off diagonal density matrix elements and hence is mostly affected by the channel \eqref{rhoBpdamp}.
The optimal measurement operator results
\begin{equation}
\hat{S}_B=\frac{1}{4(\pi^2-4)}\left(
\begin{array}{ccc}
 16-8\pi+\pi^3 & &  (\pi-4)\pi \\ \\
(\pi-4)\pi  & & 16-8\pi+\pi^3
\end{array}\right).
\end{equation}

In contrast $\bar{C}_{min}^F$ is monotonically increasing because according to \eqref{tildeKpdamp} 
the input \eqref{rhoA2qubit} is increasingly affected when changing from 
$|0\rangle\langle 0|$ to $|1\rangle\langle 1|$.

\bigskip

The privacy \eqref{Pe} can be easily evaluated by means of \eqref{CBpdamp} and \eqref{CFpdamp} as
\begin{align}\label{Pepdamp}
{\cal P}_e=\max\Bigg\{
\frac{(1-x) }{\pi ^2 (\pi^2-4 ) \left(\pi ^2 \gamma+4 \gamma+\pi ^2-4\right)}
\Big[&
\left(\pi ^6-8 \pi ^5+20 \pi ^4-32 \pi ^3+68 \pi ^2-16\right)\gamma^2\notag\\
&+\left(\pi ^6 -8 \pi ^5 +12 \pi ^4 +32 \pi ^3 -72 \pi ^2 +32\right) \gamma 
-(\pi ^4-8 \pi ^2+16)\Big],
0\Bigg\}.
\end{align}
The privacy of estimation results guaranteed only for $\gamma\in(\gamma_0,1)$, where 
$\gamma_0\approx 0.54$ (the exact expression is given in Appendix \ref{appA2}).
Furthermore the maximum is achieved for $\gamma=\gamma_{*}\approx 0.77$ (the exact expression is given in Appendix \ref{appA2}), with the measurement operator
\begin{equation}
\hat{S}_B=\frac{1}{4(\pi^2-4)}\left(
\begin{array}{ccc}
32+\pi^3-12\pi-(32 -8\pi) \gamma_* & & 4\pi(\pi-4)\sqrt{(1-\gamma_*)\gamma_*} \\ \\
4\pi(\pi-4)\sqrt{(1-\gamma_*)\gamma_*} & & \pi^3-4\pi +(32-8\pi) \gamma_*
\end{array}\right).
\end{equation}

\bigskip

As for what concern the cooperative strategy, the cost function $\bar{C}_{min}^{BF}$ obtained from Theorem \ref{P2} is shown in Fig.\ref{figpdamp}.

\begin{figure}[H]
\centering
\includegraphics[width=8cm]{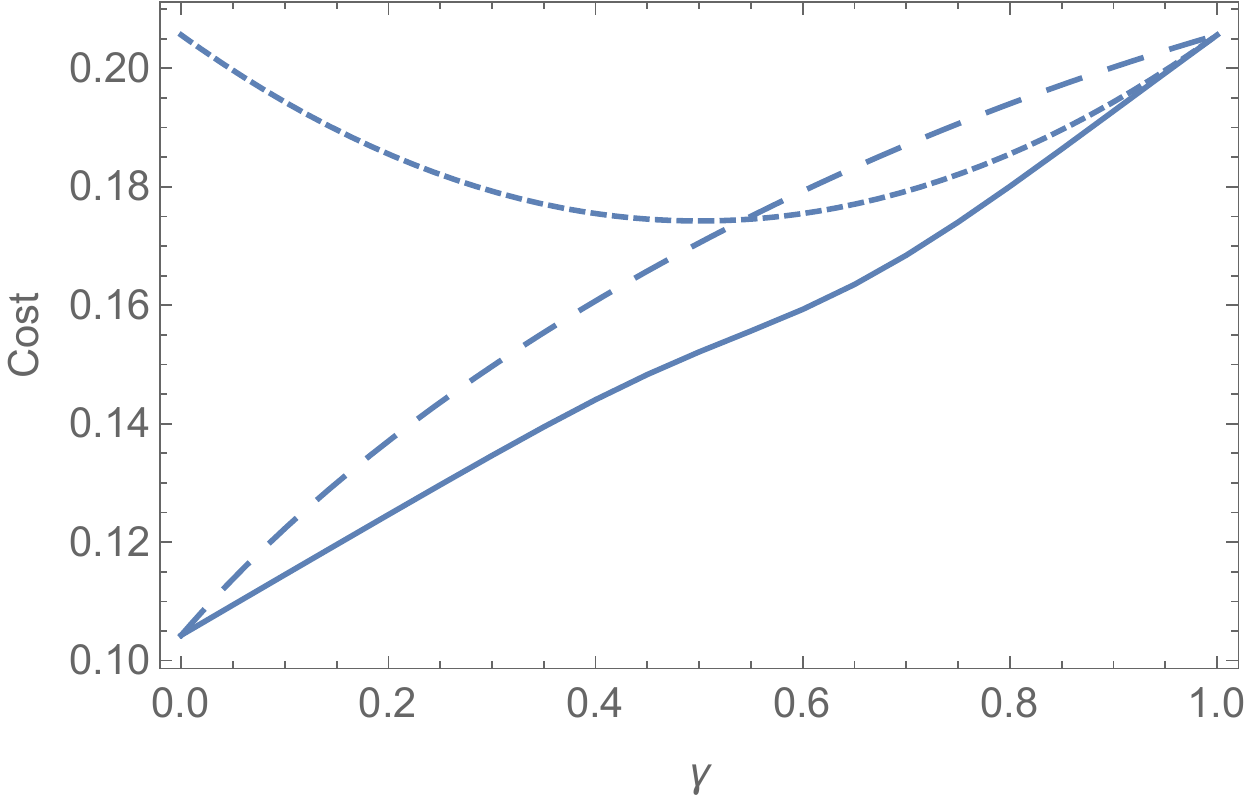}
\caption{ Plot of $\bar{C}_{min}^B$ (dotted line),
$\bar{C}_{min}^F$ (dashed line) and $\bar{C}_{min}^{BF}$ (solid line) vs $\gamma$.}
\label{figpdamp}
\end{figure}

Likewise $\bar{C}_{min}^{F}$, it reaches the minimum value $\frac{\pi^4-48}{48\pi^2}$ when 
$\gamma = 0$. The corresponding local
measurement operators read
\begin{equation}
\hat{S}_B=\left(
\begin{array}{cc}
\kappa_1 & 0 \\
0 & \kappa_2
\end{array}\right),
\qquad
\hat{S}_F=\left(
\begin{array}{cc}
\frac{\pi^2-4}{4\pi\kappa_2} & 0 \\
0 & \frac{\pi^2+4}{4\pi\kappa_2}
\end{array}\right),
\end{equation}
where $\kappa_1,\kappa_2$ are arbitrary real constant (with $\kappa_2\neq 0$).
This example shows that from Theorem \ref{P2} we can also have infinitely many solutions.

To evaluate the advantage of the cooperative strategy we consider the difference between 
the minimum average cost function of single $B$ local estimation and 
the minimum average cost function of joint $BF$ local estimation defining
\begin{equation}\label{Delta}
\Delta :=\bar{C}_{min}^{B}-\bar{C}^{BF}_{min}.
\end{equation}
From Fig.\ref{figpdamp} we may notice that the privacy cannot be guaranteed when $\Delta$ is maximum ($\gamma=0$). 
In fact in such a case the role of $F$ is dominant over $B$. It is thus reasonable to have maximum privacy away from this region, but not necessarily when $\Delta$ 
 nullifies ($\gamma=1$). In fact such a condition, although showing that the role of $F$ is irrelevant with respect to $B$, might correspond to $C^B_{min}=C^F_{min}$, which implies ${\cal P}_e=0$.

}


{
\subsection{The core set of entangling unitaries}\label{sec:app2}
}

\begin{figure}[ht]
\begin{center}
\begin{tikzpicture}[scale=0.5]
\draw[thick] (2,0) -- (2,7); 
\draw[thick,dashed] (2,0) -- (9,0.75); 
\draw[thick] (2,0) -- (5.5,-3.5); 
\draw[thick,dashed] (2,0) -- (11.5,-2.2); 
\draw[thick] (5,-3) -- (11.5,-2.2); 
\draw[thick] (11.5,4.8) -- (11.5,-2.2); 
\draw[thick] (11.5,4.8) -- (5,-3); 
\draw[thick] (11.5,4.8) -- (2,0); 
\node[left] at (2.5,8)[font = \fontsize{12}{12}]{${s_z}$}; 
\node[left] at (11,1)[font = \fontsize{12}{12}]{${s_y}$};
\node[left] at (6.5,-4.5)[font = \fontsize{12}{12}]{${s_x}$}; 
\node[left] at (2,0)[font = \fontsize{10}{10}]{$\left(0,0,0\right)$}; 
\node[left] at (5,-3.5)[font = \fontsize{10}{10}]{$\left(\frac{\pi}{2},0,0\right)$}; 
\node[left] at (13.5,-3.1)[font = \fontsize{10}{10}]{$\left(\frac{\pi}{2},\frac{\pi}{2},0\right)$}; 
\node[left] at (13.5,5.5)[font = \fontsize{10}{10}]{$\left(\frac{\pi}{2},\frac{\pi}{2},\frac{\pi}{2}\right)$}; 
\end{tikzpicture}
\caption{Tetrahedron representing the parameters space $\mathscr{S}$ of two-qubit unitaries.}
\label{tetrahedron}
\end{center}
\end{figure}
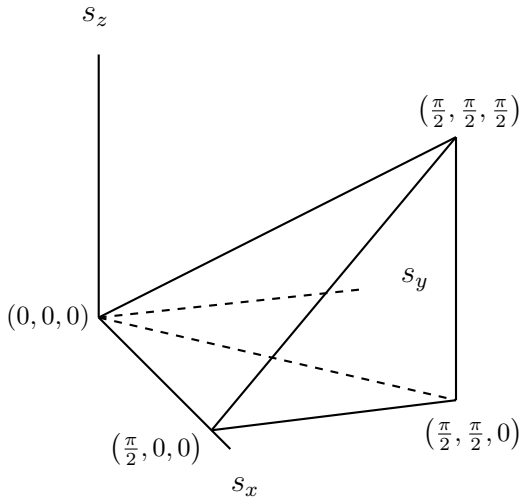

{
We now consider a set of unitaries given by}
\begin{equation}
U\left(\vec{s}\,\right)=\exp\left[-\frac{i}{2}\left(
s_x\sigma_x\otimes\sigma_x
+s_y\sigma_y\otimes\sigma_y
+s_z\sigma_z\otimes\sigma_z\right)\right],
\label{U}
\end{equation}
whose matrix representation is given in Appendix \ref{appB}. The parameter space
\begin{equation}\label{spaceS}
\mathscr{S}=\left\{\vec{s}\equiv(s_x, s_y, s_z): \frac{\pi}{2}\geq s_x \geq s_y \geq s_z \geq 0\right\},
\end{equation}
describes a tetrahedron in $\mathbb{R}^3$ as illustrated in Fig. \ref{tetrahedron}.

{
We refer to the set of unitaries of \eqref{U} as the `core set of entangling unitaries' because 
it is known that any other two-qubit entangling unitary can be traced back to this form 
by means of prior and posterior local unitaries \cite{KC01}. However these latter are not always applicable in a communication scenario, where the environment is not controllable (at beginning and/or at the 
end). Thus there are other unitaries that can be considered (an example is provided in the previous Subsection).
}

\bigskip

Below we will consider the estimation of a single parameter be either $s_x$ or $s_y$ or $s_z$ by assuming the values of the other two to be known.
While distinguishing between the two strategies described in Sections \ref{sec:aqe} and \ref{sec:cqe}, we shall also seek for optimization over probe's state, i.e. parameters $\gamma$ and $\varphi$.

\begin{rem}\label{remphi}
It can be easily checked that in the states $\rho_B$, $\rho_F$ and $\rho({\vec{s}\,})=V_{\vec{s}}\,\rho_A V_{\vec{s}}^\dag$ (see following subsections) the parameter $\varphi$ appears as added to 
${s_z}$. Thus it has no effect in the estimation of the latter.
Instead it can affect the estimation of $s_x$ and $s_y$. 
\end{rem}


\subsubsection{Adversarial quantum estimation}

The states $\rho_B$ and $\rho_F$ of Eqs.\eqref{calN} and \eqref{tildecalN},
 according to the positions \eqref{VU} and \eqref{U}, read in this case
\begin{subequations}
\begin{align}
\rho_B&={\rm Tr}_F\left[ U(\vec{s}\,) \left(\rho_A\otimes |0\rangle_E\langle 0|\right) 
U(\vec{s}\,)^\dag\right],\label{rhoB2qubit}\\
\rho_F&={\rm Tr}_B\left[ U(\vec{s}\,) \left(\rho_A\otimes |0\rangle_E\langle 0|\right) 
U(\vec{s}\,)^\dag\right],\label{rhoF2qubit}
\end{align}
\end{subequations}
where $\rho_A$ is as \eqref{rhoA2qubit}. Their matrix representation is given in Appendix \ref{appB}.

Then we distinguish the following cases:

\begin{itemize}
\item
{\bf Estimation of ${s_x}$}.

We took 325 points in the region $0\leq s_z \leq s_y \leq \frac{\pi}{2}$ 
and for each point we estimated ${s_x}$ through $\rho_B$ and independently through $\rho_F$. We actually computed 
\begin{equation}
{{\cal P}'_e}({s_y,s_z})=\max\left\{
\max_{ \gamma,\varphi} \left[\bar{C}^F_{min}(s_y,s_z,{ \gamma},\varphi)
-\bar{C}^B_{min}(s_y,s_z,{\gamma},\varphi)\right],0\right\},
\end{equation}
whose contour plot is shown in Fig.\ref{figB}.

\item
{\bf Estimation of ${s_y}$}.
 
In this case we took 325 points in the region $0\leq s_z \leq s_x \leq \frac{\pi}{2}$, and for each point we estimated ${s_y}$ through $\rho_B$ and independently through 
$\rho_F$ likewise the previous case. Then we evaluated the privacy
\begin{equation}
{ {\cal P}'_e}({s_x,s_z})=\max\left\{
\max_{ \gamma,\varphi} \left[\bar{C}^F_{min}({s_x,s_z},{\gamma},\varphi)
-\bar{C}^B_{min}({s_x,s_z},{\gamma},\varphi)\right],0\right\},
\end{equation}
whose contour plot is reported in Fig.\ref{figB}. 

Notice that although $\bar{C}^B_{min}$ can be made zero by choosing ${s_z=s_x}$\footnote{This choice by virtue of \eqref{spaceS} forces ${s_y}$ to be exactly determined.}, this does not give the maximum privacy since in such a case also $\bar{C}^F_{min}$ turns out to be zero.
Actually the maximum of privacy is obtained for ${s_z=0}$ and by increasing ${s_x}$ towards 
$\pi/2$.

\item
{\bf Estimation of ${s_z}$}.

In this last case we took 325 points in the region $ 0\leq s_y\leq s_x \leq \frac{\pi}{2}$
and for each point we estimated ${s_z}$ through $\rho_B$ and independently through 
$\rho_F$. This is done by also optimizing the privacy \eqref{Pe} over the probe's state, i.e. 
by considering 
\begin{equation}
{{\cal P}'_e}({s_x,s_y})=\max\left\{\max_{\gamma} \left[\bar{C}^F_{min}({s_x,s_y},{ \gamma})
-\bar{C}^B_{min}({s_x,s_y},{\gamma})\right],0\right\},
\end{equation}
whose contour plot is reported in Fig.\ref{figB}.

\begin{figure}[H]
\centering
\includegraphics[width=6cm]{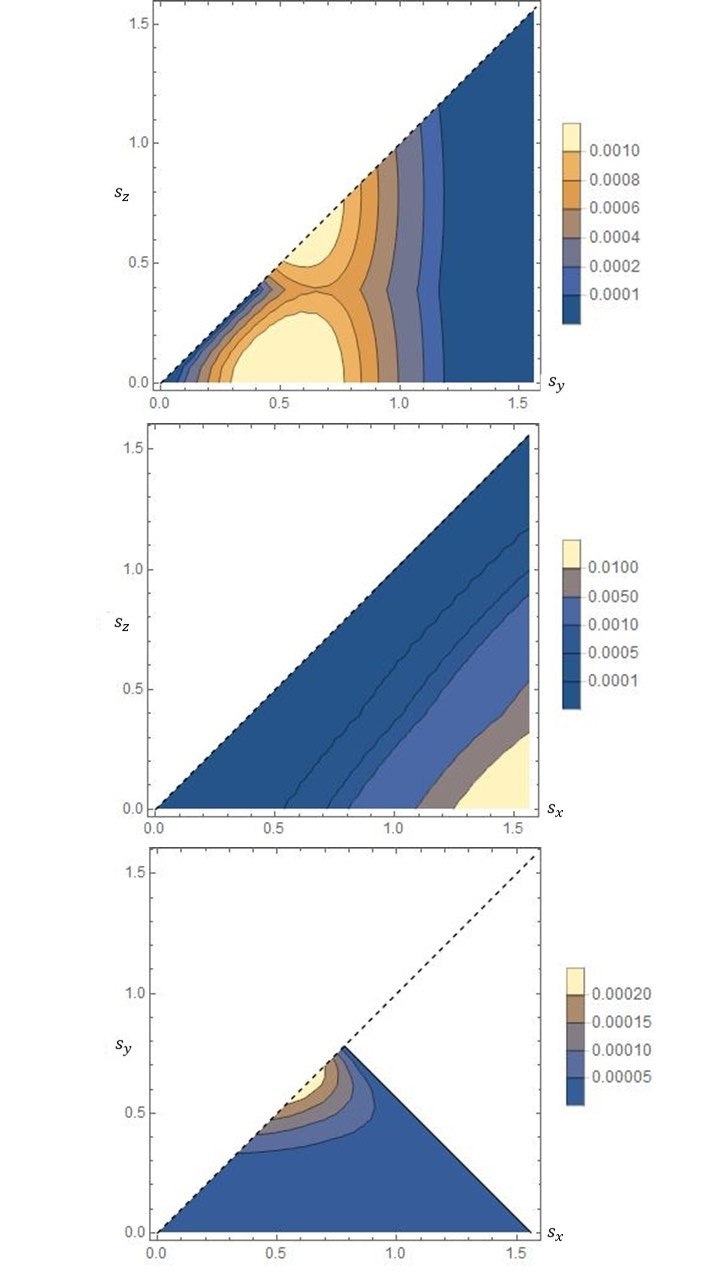}
\caption{Contour plot of the privacy ${{\cal P}'_e}$ for estimating $s_x$ (top),
$s_y$ (middle) and $s_z$ (bottom). 
The triangle above the dashed line represents the region of not admissible parameters values.}
\label{figB}
\end{figure}

On the line ${s_x+s_y}=\frac{\pi}{2}$ we have $\bar{C}^B_{min}=\bar{C}^F_{min}$ and this divides the region $ 0\leq s_y\leq s_x \leq \frac{\pi}{2}$ into two triangles. Only in the lower one the estimation is private (in the upper one $\bar{C}_{min}^F$ results smaller than $\bar{C}_{min}^B$). Furthermore there is a specific and small region where the privacy 
increases with respect to the background.

\end{itemize}

Comparing the three cases we can see that the highest privacy is achievable for the estimation of 
${s_y}$, while it decreases by one order of magnitude for ${s_x}$ and by a further order of magnitude for ${s_z}$. 
In this latter case the privacy is also not guaranteed in half of the 
parameters space. In any case the legitimate user, for a safer estimation, should set the values of other parameters in a suitable way.
It is worth saying that ${{\cal P}'_e}({s_x,s_y})$ is not affected (according to remark \ref{remphi}) by the maximization over $\varphi$, while the quantities ${{\cal P}'_e}({s_x,s_z})$ and 
${{\cal P}'_e}({s_y,s_z})$ are, but in a different way. In particular the former is almost insensible to 
$\varphi$, instead   
the latter strongly depends on it.


\subsubsection{Cooperative quantum estimation}

We start from the state of the composite system $BF$
which, according to the positions \eqref{VU} and \eqref{U}, reads
\begin{equation}
\rho(\vec{s}\,)
=\left[ U(\vec{s}\,) \left(\rho_A\otimes |0\rangle_E\langle 0| \right)
U(\vec{s}\,)^\dag\right],
\label{rhoa}
\end{equation}
where $\rho_A$ is as \eqref{rhoA2qubit}. 
Its matrix representation is reported in Appendix \ref{appB}.

We computed \eqref{W} and \eqref{WBF} in order to solve the system of nonlinear equations \eqref{1}. This has been done numerically (employing {\scshape Mathematica} packages for solving generic equations with high working precision)
for values of {  $\gamma\in[0,1]$} with step 0.1 and of $\varphi\in[0,2\pi]$ with step $\frac{\pi}{8}$ finding optimal local measurement operators.

Then we distinguish the following cases.

\begin{figure}[H]
\centering
\includegraphics[width=6cm]{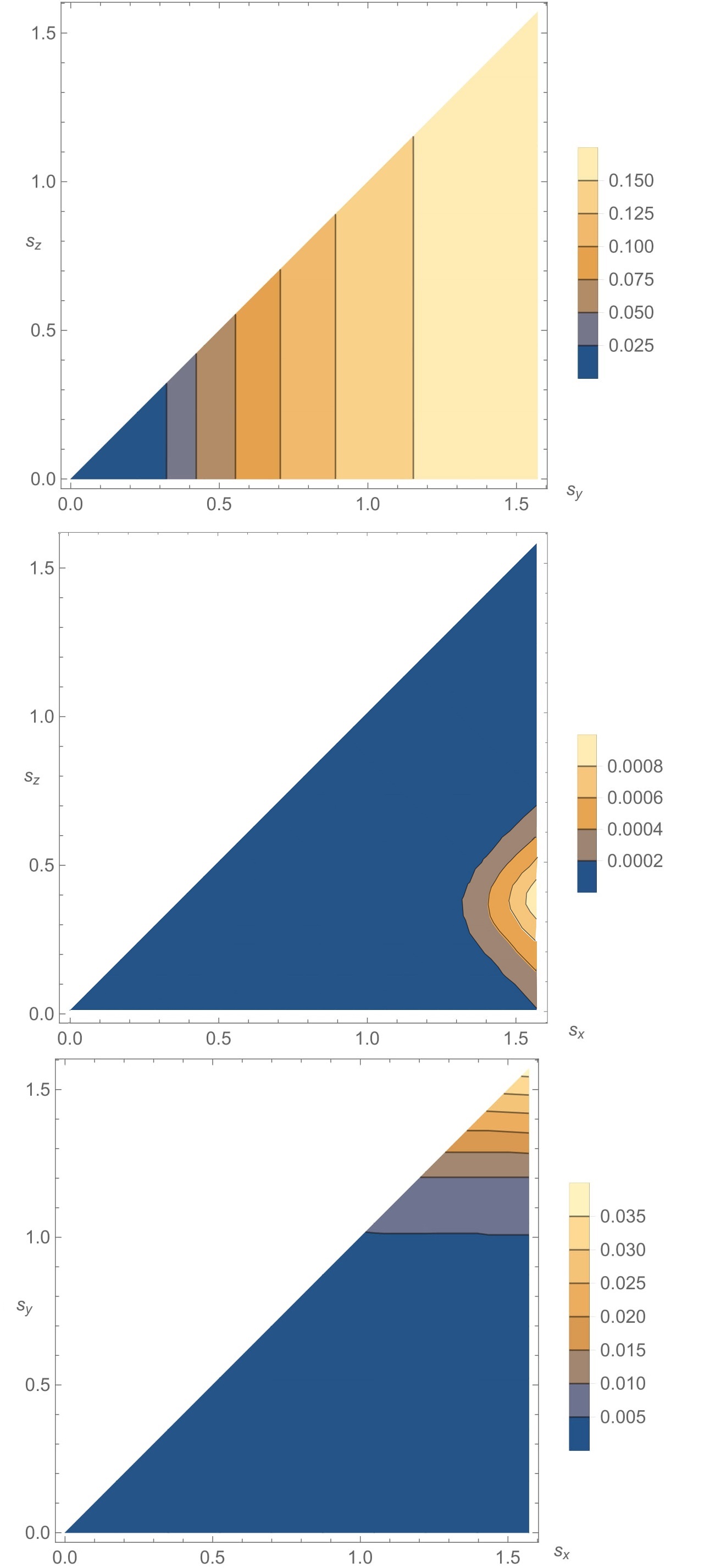}
\caption{Contour plot of the difference {  $\Delta'$
for estimating} $s_x$ (top),
$s_y$ (middle) and $s_z$ (bottom). 
The upper white triangle represents the region of not admissible parameters values.}
\label{figBF}
\end{figure}

\begin{itemize}
\item
{\bf Estimation of ${s_x}$}.

We took 325 points from the region $0\leq s_y \leq \frac{\pi}{2},\, 0\leq s_z \leq \ s_y$, and for each point evaluated \eqref{Delta} for the estimation of ${s_x}$. 
This is done by also optimizing the cost functions over the probe's state, i.e. considering 
\begin{equation}
{ \Delta'}({s_y,s_z}):=\min_{\gamma,\varphi}\left[\bar{C}^B_{min}({s_y,s_z},{ \gamma},\varphi)\right]
-\min_{  \gamma,\varphi}\left[\bar{C}^{BF}_{min}({s_y,s_z},{ \gamma},\varphi)\right],
\end{equation}
whose contour plot is shown in 
Fig.\ref{figBF}.

We can see that it does not depend on ${s_z}$.
Furthermore, it increases with ${s_y}$ and this might appear contradictory with the fact that  
at the value ${s_y=\pi/2}$ both strategies are equivalent, given that the range of estimated parameter nullifies and it can be exactly determined.
Actually this behavior is due to $\bar{C}^B_{min}$ which increases as ${s_y\to\frac{\pi}{2}}$ and then has a discontinuity in this edge (${s_y=\frac{\pi}{2}}$), where its value becomes zero. In contrast $\bar{C}^{BF}_{min}$ smoothly decreases to zero for  
${s_y\to\frac{\pi}{2}}$.

\item
{\bf Estimation of ${s_y}$}.

In this case we took 325 points from the region $ 0\leq s_x \leq \frac{\pi}{2},\, 0\leq s_z \leq \ s_x$, and for each point evaluated \eqref{Delta} for the estimation of ${s_y}$.
This is done by also optimizing the cost functions over the probe's state, i.e. considering 
\begin{equation}
{  \Delta'}({s_x,s_z}):=\min_{ \gamma,\varphi}\left[\bar{C}^B_{min}({s_x,s_z},{ \gamma},\varphi)\right]
-\min_{  \gamma,\varphi}\left[\bar{C}^{BF}_{min}({s_x,s_z},{ \gamma},\varphi)\right],
\end{equation}
whose contour plot is shown in 
Fig.\ref{figBF}.

We may notice that along the line ${s_z=s_x}$ the quantity $\Delta$ tends to zero because 
${s_y}$ becomes exactly determined. 
The major improvement due to the cooperative strategy occurs for ${s_x=\frac{\pi}{2}}$ 
{  (around ${s_z}=\frac{\pi}{8}$).
}

\item
{\bf Estimation of ${s_z}$}.

In this last case we took 325 points from the region $ 0\leq s_x \leq \frac{\pi}{2},\, 0\leq s_y \leq \ s_x$, and for each point evaluated \eqref{Delta} for the estimation of ${s_z}$.
This is done by also optimizing the cost functions over the probe's state, i.e. simply over $x$ 
according to the Remark \ref{remphi}. Hence we considered 
\begin{equation}\label{Deltaaz}
{  \Delta'}({s_x,s_y}):=\min_{ \gamma}\left[\bar{C}^B_{min}({s_x,s_y},{ \gamma})\right]
-\min_{  \gamma}\left[\bar{C}^{BF}_{min}({s_x,s_y},{ \gamma})\right],
\end{equation}
whose contour plot is shown in Fig.\ref{figBF}.

We can see no dependance on $s_x$ and, above all,
that the biggest enhancement in the estimation capability with cooperative strategy takes place on the corner ${s_x=s_y=\pi/2}$, where the parameter ${s_z}$ has the largest range. The advantage decreases towards ${s_y=0}$, where the range of ${s_z}$ reduces to zero making the estimation meaningless. 
\end{itemize}

Comparing the three cases we can see that the highest improvement due to cooperative strategy is achievable for the estimation of 
 ${s_x}$ and ${s_y}$, while it is sensibly lower for ${s_z}$. Anyway the advantage is always guaranteed in the entire parameters space.

It is worth saying that ${ \Delta'}({s_x,s_y})$ is not affected (according to remark \ref{remphi}) by the maximization over $\varphi$, while the quantities ${ \Delta'}({s_x,s_z})$ and 
${ \Delta'}({s_y,s_z})$ are only slightly affected by it (since it is the smaller quantity 
$\bar{C}^{BF}_{min}$ in the difference to be more sensible to it). 


\section{Concluding remarks}\label{sec:CR}

In summary we have considered the single parameter estimation of isometries representing Stinespring dilations of quantum channels in two different contexts. One in which the environment is under control of an adversary and the goal is to allow the legitimate user of the channels to outperform the estimation.
Another in which the environment is under control of an helper and the goal is to improve the estimation of the legitimate user of the channels. This shares analogies with feedback assistance models \cite{GW03,MCM11}, where information gathered from environment is fed back to the main system with the aim of improving the channel performance.

In both cases the optimal strategies have been found by minimizing the mean square error.
As such they are universal, i.e. not depending on the value of the estimated parameter, 
in contrast with the approach that looks for the POVM
maximizing the Fisher information, therefore minimizing the variance of the estimator,
at a fixed value of the parameter \cite{C46}.
This in the second case required a generalization of the Personik theorem \cite{Pers71} to local measurements. Such achievement has potential applications in many different contexts whenever locality constraint is imposed on quantum estimation.

{  The developed approaches have been applied to two-qubit unitaries. 
The best strategies (input and measurement) are explicitly presented for the physically relevant case of dilation of phase damping channel in Sec.\ref{sec:app1}. 
Then, the set of unitaries of Sec.\ref{sec:app2} 
shows that the largest privacy is obtainable when estimating $s_y$.
The cooperative strategy gives maximum advantage for the estimation of $s_x$. 
The results of Sec.\ref{sec:app2}, although through ${\cal P}_e'$ and $\Delta'$ support the conclusion drawn in Sec.\ref{sec:app1} for ${\cal P}_e$ and $\Delta$, namely the fact that the privacy cannot be guaranteed when cooperation benefit is maximum. 
Actually it attains its maximum away from this region, but not necessarily when 
the cooperative strategy nullifies its benefit. 

}

Clearly the private region of estimation as well as the effectiveness of local helper can depend on the structure of the unitaries, which becomes more complicate by going beyond U$(2\times 2)$. Investigations along this direction are left for future work.
In such a case it could be convenient instead of solving the nonlinear equations \eqref{1}, to randomly generate hermitiam matrices $S_F$ 
(by using e.g. Gaussian unitary ensemble \cite{Metha}) and then solve only \eqref{1a} by standard methods for Lyapunov equations. 
The minimum of $\bar{C}^B$ overall matrices $S_F$ will provide the optimal solution for cooperative strategy (notice that the case with $S_F=I$ corresponds to unassisted estimation by Bob). 

Finally, it is worth saying that the devised scenarios could be extended to 
also contemplate the action of the adversary, or the helper, on the initial state of the environment, 
rather than just on the final one. This would realize an effective channel between $A$ and $B$ \cite{sid,MM17}. 


\section*{Acknowledgments}

The work of M.R. is supported by China Scholarship Council.

\appendix
\section{Proof of Theorem \ref{P2}}\label{appA}

Being $\bar C (\hat S)$ a minimum, for $\forall H_B \in \mathscr{L}(\mathscr{H}_B)$, $H_F \in \mathscr{L}(\mathscr{H}_F)$ hermitian, 
and $\forall \epsilon_1, \epsilon_2 \in \mathbb{R}$, it must be 
\begin{subequations}
\begin{align}
\bar C (\hat S)&\leq \bar C (\hat S_B \otimes\hat S_F+\epsilon_1 H_B \otimes I 
+ \epsilon_2 I \otimes H_F) \\
&= \int_\mathscr{I} p(s) {\rm Tr} \left [\rho (s) (\hat S_B \otimes\hat S_F+\epsilon_1 H_B \otimes I + \epsilon_2 I \otimes  H_F-{s} I )^2\right ]ds \\
&= \int_\mathscr{I} p(s) {\rm Tr}\left [\rho (s) (\hat S_B \otimes\hat S_F-{s} I )^2 \right ]
ds \notag\\ 
&\quad+ \int_\mathscr{I} p(s) {\rm Tr} \left[\rho(s) (\hat S_B \otimes\hat S_F-{s} I) ( \epsilon_1 H_B \otimes I +\epsilon_2 I \otimes  H_F)\right ]ds \notag \\
&\quad+ \int_\mathscr{I} p(s) {\rm Tr}\left [\rho(s)  ( \epsilon_1 H_B \otimes I 
+ \epsilon_2 I \otimes  H_F)(\hat S_B \otimes\hat S_F-{s} I)\right ]ds \notag \\
&\quad+ \int_\mathscr{I} p(s) {\rm Tr}\left [\rho(s) ( \epsilon_1 H_B \otimes I 
+ \epsilon_2 I \otimes  H_F)^2\right ]ds. 
\label{Cepsilon12}
\end{align}
\end{subequations}
In turn, the derivatives of $\bar C (\hat S_B \otimes\hat S_F+\epsilon_1 H_B \otimes I +\epsilon_2 I \otimes H_F)$ with respect to $\epsilon_1$ and $\epsilon_2$ must be zero at $\epsilon_1=\epsilon_2=0$. Thus, using Eq.\eqref{Cepsilon12}, we get:
\begin{subequations}
\begin{eqnarray}
\frac{\partial \bar C}{\partial \epsilon_1}\Big|_{\epsilon_1=\epsilon_2=0}= \int_\mathscr{I} p(s) 
{\rm Tr}\left \{\rho(s)\left[  (\hat S_B \otimes\hat S_F-{s} I)  H_B \otimes I+ H_B \otimes I (\hat S_B \otimes\hat S_F-{s} I) )\right]\right \}ds=0,
\\ \notag \\
\frac{\partial \bar C}{\partial \epsilon_2}\Big|_{\epsilon_1=\epsilon_2=0}= \int_\mathscr{I} p(s) 
{\rm Tr}\left\{\rho(s) \left[  (\hat S_B \otimes\hat S_F-{s} I)  I \otimes H_F+ I \otimes H_F (\hat S_B \otimes\hat S_F-{s} I) )\right]\right\}ds=0.
\end{eqnarray}
\label{derivate}
\end{subequations}
Given the definitions \eqref{W}, the relations \eqref{derivate} imply
\begin{subequations}
\begin{eqnarray}
{\rm Tr}\left( H_B \otimes I \left( W^{(0)}  \left(\hat S_B \otimes\hat S_F\right) + \left(\hat S_B \otimes\hat S_F \right)  W^{(0)}-2  W^{(1)}\right)  \right) =0, \label{2a}\\ \notag \\
{\rm Tr}\left( I \otimes H_F \left( W^{(0)} \left(\hat S_B \otimes\hat S_F\right) + \left(\hat S_B \otimes\hat S_F\right)   W^{(0)}-2  W^{(1)}\right)  \right) =0. \label{2b}
\end{eqnarray}
\label{2}
\end{subequations}
Now Eq.\eqref{2a} can be rewritten as
\begin{align}
&{\rm Tr}_B\left\{{\rm Tr}_F\left( H_B \otimes I \left( W^{(0)}  \left(\hat S_B \otimes\hat S_F\right) + \left(\hat S_B \otimes\hat S_F \right)  W^{(0)}-2  W^{(1)}\right)  \right)\right\}  \notag\\
&={\rm Tr}_B\left\{ H_B {\rm Tr}_F \left( W^{(0)}  \left(\hat S_B \otimes\hat S_F\right) + \left(\hat S_B \otimes\hat S_F \right)  W^{(0)}-2  W^{(1)} \right)\right\} \notag\\
&={\rm Tr}_B\left\{ H_B  \left( {\widetilde W}_B^{(0)} \hat S_B 
+ \hat S_B {\widetilde W}_B^{(0)}-2  W^{(1)}_B \right)\right\} 
=0. 
\label{4}
\end{align}
The last line can be seen as the Hilbert-Schmidt scalar product in $\mathscr{L}(\mathscr{H}_B)$
between $H_B$ and $\left( {\widetilde W}_B^{(0)} \hat S_B 
+ \hat S_B {\widetilde W}_B^{(0)}-2  W^{(1)}_B \right)$.
Given the arbitrariness of $H_B$ we may conclude that it must be  
\begin{equation}
 {\widetilde W}^{(0)}_B \hat S_B 
+ \hat S_B {\widetilde W}^{(0)}_B=2  W^{(1)}_B.
\end{equation}
With same reasoning from Eq.\eqref{2b} we can get
\begin{equation}
 {\widetilde W}^{(0)}_F \hat S_F 
+ \hat S_F {\widetilde W}^{(0)}_F=2  W^{(1)}_F. 
\end{equation}


{ 
\section{Coefficients $\gamma_0$ and $\gamma_*$}\label{appA2}

Equating ${\cal P}_e$ of \eqref{Pepdamp} to zero we get:
\begin{equation}
\gamma_0=\left(\pi ^2-4\right)\frac{\pi\sqrt{\pi ^2 (\pi^2 -8\pi+20) (\pi -4)^2+16}-(\pi -4)^2 \pi^2+8}{2 \pi ^2 (\pi^4-8\pi^3+20\pi^2-32\pi+68)-32}.
\end{equation}
Still referring to \eqref{Pepdamp}, solving $d{\cal P}_e/d\gamma=0$ with respect to $\gamma$ yields:
\begin{equation}
\gamma_*=\frac{\Theta^2-2^{1/3}b\Theta+2^{2/3}(b^2-3ac)}{2^{1/3}3a\Theta},
\end{equation}
where 
\begin{equation}
\begin{split}
a&:=2 \left(\pi ^2+4\right) \left(-16+68 \pi ^2-32 \pi ^3+20 \pi ^4-8 \pi ^5+\pi ^6\right),\\
b&:=384-1376 \pi ^2+640 \pi ^3-208 \pi ^4+64 \pi ^5+40 \pi ^6-24 \pi ^7+3 \pi ^8,\\
c&:=8(\pi^2 -4) \left(12-35 \pi ^2+16 \pi ^3-2 \pi ^4\right),\\
d&:=(\pi^2 -4)^2 \left(8-18 \pi ^2+8 \pi ^3-\pi ^4\right),\\
\Theta&:=\sqrt[3]{\sqrt{\left(27 a^2 d-9 a b c+2 b^3\right)^2-4 \left(b^2-3 a c\right)^3}-27 a^2 d+9 a b c-2 b^3}.
\end{split}
\end{equation}
}


\section{Operators Matrix Representation}\label{appB}

Matrix representation of Eq.\eqref{U} (non null elements) in the canonical basis $\{\ket{0}\ket{0},\ket{0}\ket{1}, \ket{1}\ket{0},\ket{1}\ket{1}\}$:
\begin{equation}
\begin{split}
[U(\vec{s}\,)]_{11}&=
 e^{-\frac{1}{2} i s_z} \cos \left(\frac{s_x-s_y}{2}\right), \\
[U(\vec{s}\,)]_{14}&= -i e^{-\frac{1}{2} i s_z} \sin \left(\frac{s_x-s_y}{2}\right), \\
[U(\vec{s}\,)]_{22}&= e^{\frac{i }{2}s_z} \cos \left(\frac{s_x+s_y}{2}\right), \\
[U(\vec{s}\,)]_{23} &=-i e^{\frac{i }{2}s_z} \sin \left(\frac{s_x+s_y}{2}\right), \\
 [U(\vec{s}\,)]_{32}&= -i e^{\frac{i }{2}s_z} \sin \left(\frac{s_x+s_y}{2}\right), \\
 [U(\vec{s}\,)]_{33}&= e^{\frac{i }{2}s_z} \cos \left(\frac{s_x+s_y}{2}\right), \\
[U(\vec{s}\,)]_{41}&= -i e^{-\frac{1}{2} i s_z} \sin \left(\frac{s_x-s_y}{2}\right), \\
[U(\vec{s}\,)]_{44}&=e^{-\frac{1}{2} i s_z} \cos \left(\frac{s_x-s_y}{2}\right).
\end{split}
\end{equation}


Matrix representation of Eq.\eqref{rhoB2qubit} in the canonical basis $\{\ket{0},\ket{1}\}$:
\begin{equation}
\begin{split}
[\rho_B]_{11}&=
 \frac{1}{2}+\left(x-\frac{1}{2}\right) \cos s_x \cos s_y+\frac{1}{2}\sin s_x \sin s_y, \\
[\rho_B]_{12}&= \sqrt{(1-x) x} \left(\cos s_y \cos (s_z+\varphi)-i \cos s_x \sin (s_z+\varphi)\right), \\
[\rho_B]_{22}&= \frac{1}{2}-\left(x-\frac{1}{2}\right) \cos s_x \cos s_y-\frac{1}{2}\sin s_x \sin s_y.
\end{split}
\end{equation}


Matrix representation of Eq.\eqref{rhoF2qubit} in the canonical basis $\{\ket{0},\ket{1}\}$:
\begin{equation}
\begin{split}
[\rho_F]_{11}&=
  \frac{1}{2}+\left(x- \frac{1}{2}\right) \sin s_x \sin s_y+ \frac{1}{2}\cos s_x \cos s_y, \\
[\rho_F]_{12}&= \sqrt{(1-x) x} \left( \sin s_y \sin (s_z+\varphi) + i  \sin s_x \cos (s_z+\varphi) \right), \\ 
[\rho_F]_{22}&=  \frac{1}{2}-\left(x- \frac{1}{2}\right) \sin s_x \sin s_y
 - \frac{1}{2}\cos s_x \cos s_y.
\end{split}
\end{equation}


Matrix representation of Eq.\eqref{rhoa} in the canonical basis $\{\ket{0}\ket{0},\ket{0}\ket{1}, \ket{1}\ket{0},\ket{1}\ket{1}\}$:
\begin{equation}
\begin{split}
[\rho(\vec{s}\,)]_{11}&=\frac{1}{2}x\left(1+\cos s_x\cos s_y+\sin s_x\sin s_y\right),\\
[\rho(\vec{s}\,)]_{12}&=\frac{i}{2}\sqrt{x(1-x)}\left(\sin s_x+\sin s_y\right)e^{-i(s_z+\varphi)},\\
[\rho(\vec{s}\,)]_{13}&=\frac{1}{2}\sqrt{x(1-x)}\left(\cos s_x+\cos s_y\right)e^{-i(s_z+\varphi)},\\
[\rho(\vec{s}\,)]_{14}&=\frac{i}{2}x\left(\sin s_x\cos s_y-\cos s_x\sin s_y\right),\\
[\rho(\vec{s}\,)]_{22}&=\frac{1}{2}(1-x)\left(1-\cos s_x\cos s_y+\sin s_x\sin s_y\right),\\
[\rho(\vec{s}\,)]_{23}&=-\frac{i}{2}(1-x)\left(\cos s_x\sin s_y+\sin s_x\cos s_y\right),\\
[\rho(\vec{s}\,)]_{24}&=-\frac{1}{2}\sqrt{x(1-x)}\left(\cos s_x-\cos s_y\right)
e^{i(s_z+\varphi)},\\
[\rho(\vec{s}\,)]_{33}&=\frac{1}{2}(1-x)\left(1+\cos s_x\cos s_y-\sin s_x\sin s_y\right),\\
[\rho(\vec{s}\,)]_{34}&=\frac{i}{2}\sqrt{x(1-x)}\left(\sin s_x-\sin s_y\right)
e^{i(s_z+\varphi)},\\
[\rho(\vec{s}\,)]_{44}&=\frac{1}{2}x\left(1-\cos s_x\cos s_y-\sin s_x\sin s_y\right).
\end{split}
\end{equation}


\end{document}